\documentclass{article}
\usepackage{spconf,amsmath,graphicx}
\usepackage{booktabs,multirow}
\usepackage{amsfonts}
\usepackage{xcolor}

\usepackage{enumitem}
\setlist{nosep, leftmargin=14pt}

\usepackage{mwe} 

\title{BrainTAP: Brain Disorder Prediction with Adaptive \\ Distill and Selective Prior Integration}
\name{Zhenyu Lei$^{\star}$, Aiying Zhang$^{\star}$, Song Wang$^{\dagger}$, Han Fan$^{\star}$, Jundong Li$^{\star}$}
\address{$^{\star}$University of Virginia 
    $^{\dagger}$University of Central Florida}
\begin{document}
%
\maketitle
\begin{abstract}
Predicting clinical outcomes from brain networks in large-scale neuroimaging cohorts such as the Adolescent Brain Cognitive Development (ABCD) study requires effectively integrating functional connectivity (FC) and structural connectivity (SC) while incorporating expert neurobiological knowledge. However, existing multimodal fusion approaches are shallow or over-homogenize the inherently heterogeneous characteristics of FC and SC, while expert-defined anatomical priors are underutilized with static integration. To address these limitations, we propose Brain Transformer with Adaptive Mutual-Distill and Selective Prior Fusion (BrainTAP). We introduce Adaptive Mutual Distill (AMD), which enables layer-wise information exchange between modalities through learnable distill-intact ratios, preserving modality-specific signals while capturing cross-modal synergies. We further develop Selective Prior Fusion (SPF), which integrates expert-defined anatomical priors in an adaptive way. Evaluated on the ABCD dataset for predicting attention-related disorders, BrainTAP achieves superior performance over state-of-the-art baselines, demonstrating its effectiveness for brain disorder prediction.
\end{abstract}
\section{Introduction}
\label{sec:intro}
Brain connectivity networks derived from neuroimaging provide meaningful representations of brain organization linked to cognition and mental health~\cite{park2013structural}. Functional connectivity (FC) reflects correlations of blood-oxygen-level-dependent (BOLD) signal fluctuations between brain regions during resting-state functional magnetic resonance imaging (fMRI), capturing dynamic functional interactions. Structural connectivity (SC) represents anatomical white matter pathway strengths between regions estimated from diffusion magnetic resonance imaging (dMRI) tractography, reflecting the brain's physical architecture. Recent studies have increasingly used these networks to predict brain disorders. Graph-based methods such as BrainGNN apply graph convolutions to FC networks~\cite{li2021braingnn}, whereas transformer-based models like the Brain Network Transformer leverage attention mechanisms to learn pairwise connection strengths~\cite{kan2022brain}. However, most existing approaches rely on a single modality, overlooking the complementary information from FC and SC.

To unlock the complementary strengths of FC and SC, researchers have turned to multimodal fusion frameworks. Early approaches took a shallow path which simply concatenate or average FC and SC features~\cite{abrol2019multimodal} or treating them as separate “views” in multi-view embedding learning frameworks~\cite{liu2018multi}, which failing to capture deeper and more intricate synergism between these two modalities. 
To enhance the interaction between the two modalities, more advanced approaches have been developed. These include hybrid adjacency matrices that integrate FC and SC~\cite{zhang2021deep}, cross-attention mechanisms that enable each modality to directly attend to the other~\cite{khajehnejad2025brainsymphony}, and multi-scale fusion strategies applied not only at the whole-brain graph level but also at the regional scale~\cite{liu2025predicting}, leading to deeper integration and improved prediction. However, most deep-fusion frameworks can inadvertently homogenize the inherently heterogeneous FC and SC information by over-emphasizing their integration, thereby impairing modality-specific features critical for prediction.
Beyond methodological issues, most models rely solely on imaging data while neglecting the value of expert-derived neurobiological priors, such as established knowledge about connectivity patterns already linked to particular cognitive functions or clinical outcomes. Such priors can guide model focus, improve interpretability, and ground predictions in anatomically meaningful ways. Yet when incorporated, these priors are commonly imposed as static binary masks that treat all prior regions as equally informative~\cite{brown2018connectome}. This coarse constraint overlooks the substantial variability in regional contributions, and effective integration requires fine-grained, region-specific importance weights.


To address these challenges, we propose a novel framework termed \textbf{Brain Transformer with Adaptive Mutual-Distill and Selective Prior Fusion (BrainTAP)}. 
For the first challenge, we introduce \textbf{Adaptive Mutual Distill (AMD)}, a mechanism in which the two modalities progressively exchange information through layer-wise mutual distillation. At each layer, only a fraction of features are distilled while the rest remain modality-intact, preventing over-homogenization. To achieve this, AMD learns a distill–intact ratio that dynamically controls the extent of sharing across layers. This adaptive design enables the model to distill cross-modal knowledge where beneficial, while preserving modality-specific signals that are essential for capturing heterogeneous FC–SC characteristics.
For the second challenge, we propose \textbf{Selective Prior Fusion (SPF)}, a framework that integrates expert-defined anatomical priors in a selective way. Rather than imposing static binary masks uniformly across all regions, SPF learns region-specific masks that modulate the influence of neurobiological priors. These masks comprise two complementary components: a global-level component that encodes population-level neurobiological importance shared across subjects, and an individual-level component conditioned on subject-specific embeddings. Together, these components generate a selective attention mask that adaptively modulates connectivity importance within the Transformer architecture for each subject.
Empirical results on the ABCD~\cite{hagler2019image} dataset show that BrainTAP achieves superior performance in predicting multiple attention-related disorders, surpassing state-of-the-art baselines and validating the effectiveness of our proposed framework.

\begin{figure}[t]
    \centering
    \includegraphics[width=\linewidth]{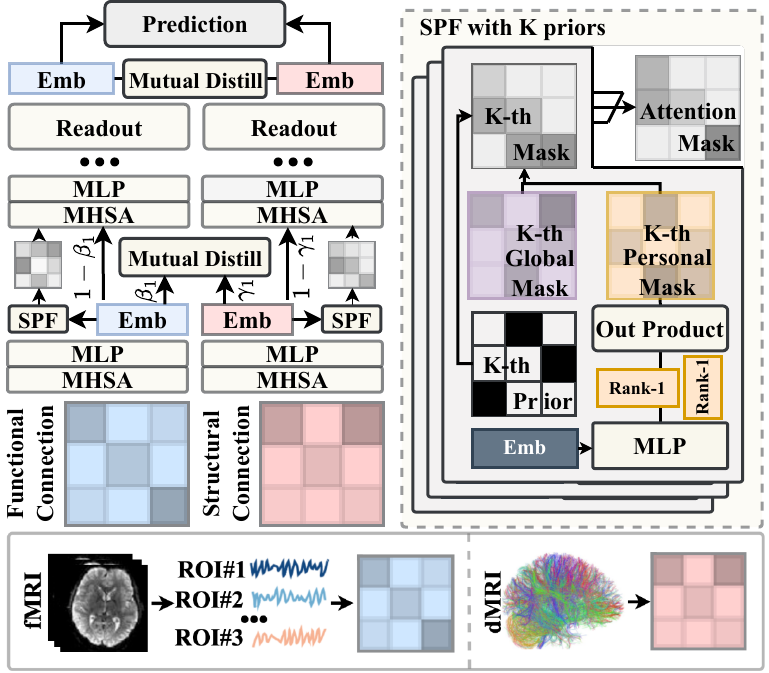}
    \caption{Overview of BrainTAP framework. FC and SC are represented as blue and pink adjacency matrices, respectively.}
    \label{fig:overview}
\end{figure}

\section{Methodology}

An overview of BrainTAP is presented in Figure~\ref{fig:overview}. The framework builds upon the Transformer architecture to capture nuanced brain connections and comprises two modules: (1) Adaptive Mutual Distillation (AMD), which enables controlled cross-modal information exchange through layer-wise mutual distillation while preserving modality-specific signals, and (2) Selective Prior Fusion (SPF), which selectively integrates anatomical priors with learnable global and personalized prior masks. 

\vspace{5pt}
\noindent\textit{\textbf{Problem Formulation.}} 
For each subject, we observe two weighted brain networks over $N$ ROIs including FC as $A^{F}\!\in\!\mathbb{R}^{N\times N}$ and SC as $A^{S}\!\in\!\mathbb{R}^{N\times N}$. Our objective is to predict disease status $y \in \{0,1\}$ given these two networks as input.

\subsection{Transformer Architecture}
For modality indicator $M\!\in\!\{F,S\}$, the connectivity fingerprint of ROI $i$ is the $i$-th row $\mathbf{a}^{M}_i$ of $A^{M}$, embedded by a modality-specific map $\phi^{M}$ to tokens $X^{M,0}$ with rows $\mathbf{x}^{M,0}_i=\phi^{M}(\mathbf{a}^{M}_i)$ utilized as the input feature of ROI $i$. Each modality is encoded by an $L$-layer Transformer with multi-head self-attention (MHSA) and an MLP:
\[
P^{M,\ell}_h=\mathrm{softmax}\!\big(Q_hK_h^\top/\sqrt{d}\big),
\]
\[
\mathrm{MHSA}(X)=\mathrm{Concat}_h\!\big(P^{M,\ell}_h V_h\big)W_O,
\]
\[
Z^{M,\ell}=X^{M,\ell-1}+\mathrm{MHSA}(X^{M,\ell-1}),
\]
\[
X^{M,\ell}=Z^{M,\ell}+\mathrm{MLP}(Z^{M,\ell}),
\]
where $Q_h=XW_{Q_h}$, $K_h=XW_{K_h}$, $V_h=XW_{V_h}$, $d$ is the per-head dimension, and $\ell=1,\dots,L$. The final representations $X^{F,L}$ and $X^{S,L}$ are aggregated via mean pooling over the $N$ ROIs, averaged, and passed through an MLP classifier to produce the disease prediction $\hat{y}$.

\subsection{Adaptive Mutual Distillation}
To encourage complementary yet aligned representations, we distill information between modalities at each layer and inject it back as a residual connection. Let $g_F, g_S$ denote MLP projection heads that map into a shared $d$-dimensional distillation space, and $h_F, h_S$ denote their corresponding inverse projections. From the post-MLP representations, we compute
\[
Z^{F,\ell} = g_F(X^{F,\ell}), \qquad
Z^{S,\ell} = g_S(X^{S,\ell}).
\]
For each ROI $i$, we define softened channel-wise distributions
\[
p_i^{F,\ell} = \mathrm{softmax}(Z^{F,\ell}_{i,:}/\tau), \qquad
p_i^{S,\ell} = \mathrm{softmax}(Z^{S,\ell}_{i,:}/\tau),
\]
where $\tau > 0$ is a temperature parameter. A symmetric mutual distillation loss aligns the distributions from both modalities:
\[
\mathcal{L}_{\mathrm{distill}}^{\ell}
= \frac{1}{N}\sum_{i=1}^{N}[
\beta^\ell \,\mathrm{KL}(p_i^{F,\ell} \,\|\, p_i^{S,\ell})
+ \gamma^\ell \,\mathrm{KL}(p_i^{S,\ell} \,\|\, p_i^{F,\ell})].
\]
The distilled information is injected across modalities via residual connections:
\[
\Delta X^{F,\ell} = \gamma^\ell\, h_F(Z^{S,\ell}), \qquad
\Delta X^{S,\ell} = \beta^\ell\, h_S(Z^{F,\ell}),
\]
yielding updated token representations
\[
\widetilde{X}^{F,\ell} = (1-\gamma^\ell) X^{F,\ell} + \Delta X^{F,\ell},
\widetilde{X}^{S,\ell} = (1-\beta^\ell) X^{S,\ell} + \Delta X^{S,\ell},
\]
where learnable $\beta^\ell, \gamma^\ell \in [0,1]$ control the distillation strength for the SC-to-FC and FC-to-SC transfers, respectively. These refined representations $\widetilde{X}^{M,\ell}$ serve as inputs to layer $\ell + 1$.

\subsection{Selective Prior Fusion}
Anatomical priors derived from expert knowledge provide valuable structural constraints, yet simply imposing them as static binary masks uniformly is suboptimal. We propose \textbf{Selective Prior Fusion (SPF)}, which learns shared population-level masks and subject-specific individual-level masks to achieve both adaptivity and interpretability.

We begin with $K$ binary anatomical prior masks $\{\Pi_k\in\{0,1\}^{N\times N}\}_{k=1}^K$ that encode expert-defined critical regions. To ensure complete coverage, we define a complementary region $\Pi_{\mathrm{free}}=\mathbf{1}-\max_{k}\Pi_k$ for connections not assigned to any prior, yielding the complete set $\mathcal{P}=\{\Pi_1,\dots,\Pi_K,\Pi_{\mathrm{free}}\}$.

For each prior $\Pi_k$, we learn two components: a global mask matrix $W_k^{(g)}\in\mathbb{R}^{N\times N}$ shared across all subjects, and a personalized rank-1 mask matrix that adapts to individual variations. Given a learned subject embedding $\mathbf{z}$, the personalized component is generated as
\[
W_k^{(r)}(\mathbf{z})=\frac{\alpha_k(\mathbf{z})}{2}\Big[u_{k,A}(\mathbf{z})u_{k,B}(\mathbf{z})^\top+u_{k,B}(\mathbf{z})u_{k,A}(\mathbf{z})^\top\Big],
\]
where an MLP predicts the scalar weight $\alpha_k(\mathbf{z})\in\mathbb{R}$ and rank-1 factors $u_{k,A}(\mathbf{z}), u_{k,B}(\mathbf{z})\in\mathbb{R}^N$ from $\mathbf{z}$. This low-rank factorization enables efficient personalized mask learning.

We then aggregate contributions from all priors to form a unified importance score matrix:
\[
S(\mathbf{z})=\mathrm{sym0}(\sum_{k\in\mathcal{P}}(W_k^{(g)}+W_k^{(r)}(\mathbf{z}))\odot \Pi_k),
\]
where $\mathrm{sym0}(M)=\frac{1}{2}(M+M^\top)-\operatorname{diag}(\operatorname{diag}(M))$ enforces symmetry and removes self-connections. The resulting score matrix $S(\mathbf{z})$ reflects the predicted importance of each ROI pair for the given subject. These scores are converted into soft, differentiable gates through a sigmoid function:
\[
G(\mathbf{z})=\sigma\!({S(\mathbf{z})}/{\tau})\in(0,1)^{N\times N},
\]
where $\tau>0$ controls the sharpness of the gating mechanism.
The learned gates $G(\mathbf{z})$ are integrated into the model as attention biases, augmenting the Transformer attention logits as $QK^\top/\sqrt{d}\mapsto QK^\top/\sqrt{d}+\eta G(\mathbf{z})$, where $\eta$ scales the bias term. Through this design, SPF selectively integrates anatomical knowledge by focusing only on truly important regions.

\begin{table}[t]
\centering
\small
\resizebox{0.95\linewidth}{!}{
\begin{tabular}{lcccc}
\toprule
\textbf{Model} & \textbf{ADHD} & \textbf{Anxiety} & \textbf{OCD} & \textbf{ATT} \\
\midrule
MLP        & $55.45\pm0.90$ & $57.62\pm1.22$ & $54.38\pm0.63$ & $62.12\pm0.15$ \\
MaskedGCN  & $58.15\pm2.59$ & $66.19\pm2.54$ & $57.17\pm0.63$ & $\underline{67.40}\pm2.27$ \\
RH-BrainFS & $\underline{60.10}\pm0.50$ & $61.44\pm1.61$ & $57.88\pm0.11$ & $64.40\pm0.01$ \\
CrossGNN   & $57.95\pm0.20$ & $\underline{67.25}\pm0.62$ & $\underline{60.15}\pm0.13$ & $62.63\pm0.10$ \\
\textbf{BrainTAP}       & $\textbf{61.25}\pm2.03$ & $\textbf{73.08}\pm2.02$ & $\textbf{61.67}\pm2.58$ & $\textbf{73.77}\pm0.59$ \\
\bottomrule
\end{tabular}
}
\caption{Main results (AUC, 3 random seeds). The best and second-best scores are \textbf{bold} and \underline{underlined}.}
\label{tab:main}
\end{table}

\section{Experiment}
In this section, we aim to address five research questions: \textbf{RQ1}: How does BrainTAP perform compared to existing baselines? \textbf{RQ2}: What is the contribution of each BrainTAP component? \textbf{RQ3}: How does AMD adaptively regulate modality interaction strength? \textbf{RQ4}: How does SPF leverage prior knowledge? \textbf{RQ5}: How does BrainTAP identify critical ROIs for prediction?

\vspace{-10pt}
\subsection{Experimental Settings}
\noindent\textbf{Dataset.}
This study uses resting-state fMRI (rs-fMRI) data from the ABCD Study. Of the 11,099 participants aged 9–11 enrolled at baseline, 7,844 remained after excluding those without usable rs-fMRI scans or who failed quality control.
Data were preprocessed with the standardized ABCD pipelines~\cite{hagler2019image}. Cortical regions were parcellated with the Glasser atlas and subcortical regions with the Aseg atlas. Functional connectivity was estimated as the statistical association between region-of-interest (ROI) time series.
Our analyses focus on predicting three pediatric psychiatric conditions: obsessive–compulsive disorder (OCD), anxiety, attention-deficit/hyperactivity disorder (ADHD), and general attention disorder (ATT). Four prior masks are provided: \textit{Updating} linked to working-memory revision; \textit{Performance} indexing error monitoring and performance adjustment; \textit{Inhibition} supporting response suppression; and \textit{Attention} mediating top-down allocation and sustained focus.

\noindent\textbf{Baseline.} The comparison includes two \textbf{single-modal} baselines (1) \textit{MLP} and (2) \textit{MaskedGCN} and two \textbf{multi-modal} baselines (3) \textit{RH-BrainFS}~\cite{ye2023rh} and (4) \textit{CrossGNN}~\cite{yang2023mapping}. For the single-modal baselines, we apply each model separately to FC and SC, and average the last-layer hidden states.

\noindent\textbf{Implementation Details.} We use AUC as the evaluation metric. We set the learning rate to $1\times 10^{-4}$, weight decay to $1\times 10^{-6}$, and batch size to $32$. The temperature parameter $\tau$ is fixed at $1$, the hidden dimension $d$ is set to $64$, and the number of layers is set to $3$.

\vspace{-10pt}
\subsection{Main Results}
To address \textbf{RQ1}, we provide the results in Table~\ref{tab:main} and observe that: (1) BrainTAP consistently outperforms all baselines, confirming its effectiveness in predicting different brain disorders. (2) Multi-modal baselines surpass single-modal ones, highlighting the value of interaction of FC and SC representations. (3) BrainTAP performs better for anxiety and ATT, indicating that these disease may have a larger impact on brain structure and function and the abnormal connectivity is easier to capture. To examine \textbf{RQ2}, we perform an ablation study to quantify the contribution of each component. We use “\textit{w/o}” to indicate removal; “\textit{G-SPF}” and “\textit{P-SPF}” denote the global and personalized SPF variants, respectively. As shown in Table~\ref{tab:ablation}, removing any component degrades performance, confirming the effectiveness of all components.

\begin{figure}
    \centering
    \includegraphics[width=0.9\linewidth]{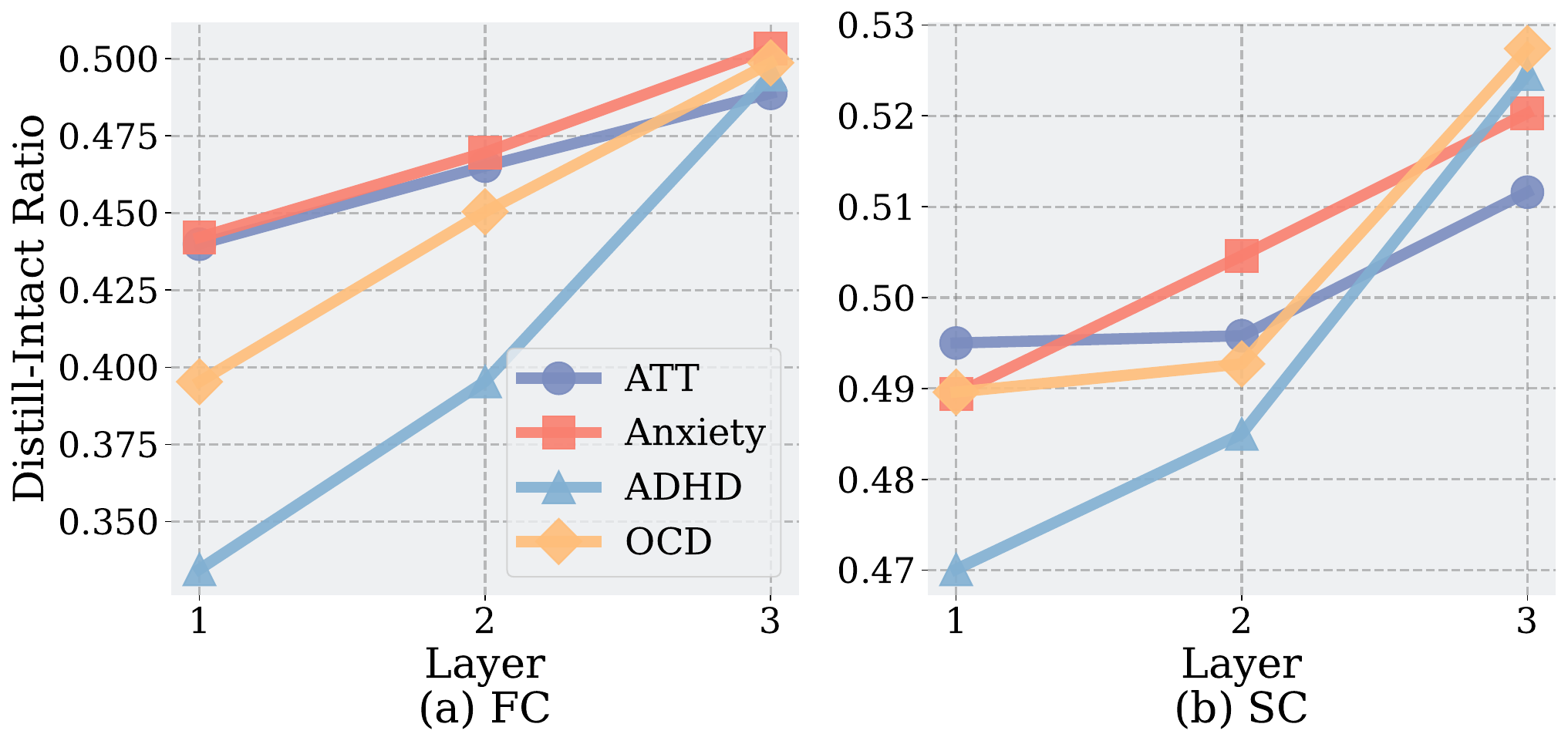}
    \caption{Distill–intact ratio across AMD layers for FC and SC. Larger values indicate stronger cross-modal interaction.}
    \label{fig:distill_ratio}
\end{figure}


\begin{table}[t]
\centering
\small
\resizebox{0.95\linewidth}{!}{
\begin{tabular}{lcccc}
\toprule
\textbf{Task} & \textbf{w/o AMD} & \textbf{w/o G-SPF} & \textbf{w/o P-SPF} & \textbf{BrainTAP} \\
\midrule
ADHD   & $59.74\pm0.50$ & $59.03\pm1.26$ & $\underline{59.77}\pm0.36$ & $\textbf{61.25}\pm2.03$  \\
Anxiety& $69.19\pm2.57$ & $69.59\pm0.18$ & $\underline{71.15}\pm1.24$ & $\textbf{73.08}\pm2.02$ \\
OCD    & $59.21\pm0.65$ & $58.34\pm2.24$ & $\underline{60.19}\pm0.27$ & $\textbf{61.67}\pm2.58$ \\
ATT    & $70.40\pm1.07$ & $67.65\pm2.12$ & $\underline{71.40}\pm1.86$ & $\textbf{73.77}\pm0.59$ \\
\bottomrule
\end{tabular}}
\caption{Ablation study results (AUC, 3 random seeds). Best and second-best scores are in \textbf{bold} and \underline{underlined}.}
\label{tab:ablation}
\end{table}

\subsection{Further Analysis}
To address \textbf{RQ3}, we compute the distill–intact ratios from the AMD module. As shown in Figure~\ref{fig:distill_ratio}, the ratio increases with depth for both FC and SC, indicating that early layers primarily retain modality-specific information while deeper layers progressively integrate cross-modal information. These results demonstrate that AMD effectively enables adaptive control over inter-modal information exchange.

\begin{figure}
    \centering
    \includegraphics[width=0.95\linewidth]{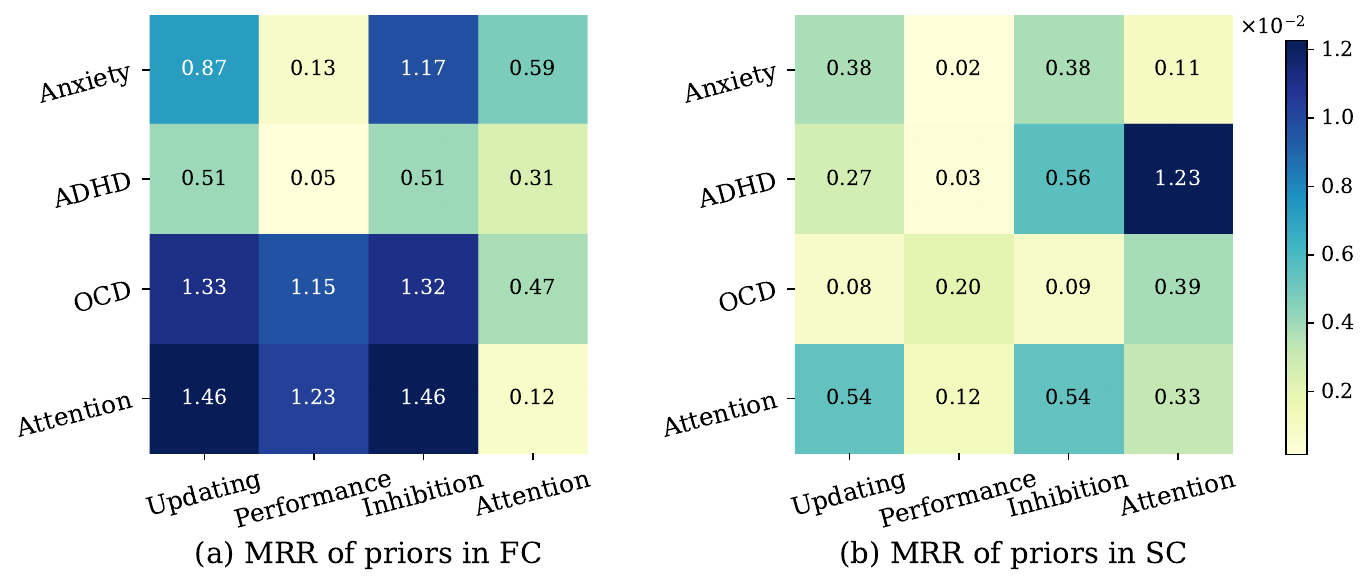}
    \caption{MRR of priors under the learned attention mask. Higher MRR indicates higher prior-disease association.}
    \label{fig:heatmap}
\end{figure}

To address \textbf{RQ4}, we compute the mean reciprocal rank (MRR) of prior regions under the learned attention mask. We first rank prior regions by attention mask values, compute the subject-level MRR, then average across subjects to obtain the prior–disease association score. As shown in Figure~\ref{fig:heatmap}, connectivity within cognitive-control circuits -- particularly those supporting goal updating and response inhibition -- exhibited strong associations across all the diseases, suggesting that disruptions in executive processes may underlie shared vulnerabilities across these conditions. This aligns with evidence that the frontoparietal control network (FPN) and cingulo-opercular/salience systems (CON/SN) are central to cognitive control and dysregulated across these disorders \cite{Cole2014Frontoparietal}. Specifically, ADHD studies report altered FPN and fronto-striatal connectivity linked to impaired inhibition and attentional control \cite{Konrad2010adhd}; OCD studies show aberrant prefrontal, orbitofrontal, and cingulo-opercular activity during inhibitory control and error processing \cite{Norman2018Error}; and comparative work highlights overlapping deficits in networks supporting interference control and performance monitoring \cite{brem2014neurobiological}. The attention construct was not prominently featured in our results in FC, likely because we used resting-state fMRI, where attention networks are less active than during externally-oriented tasks \cite{Seitzman2019}.  

\begin{figure}
    \centering
    \includegraphics[width=0.9\linewidth]{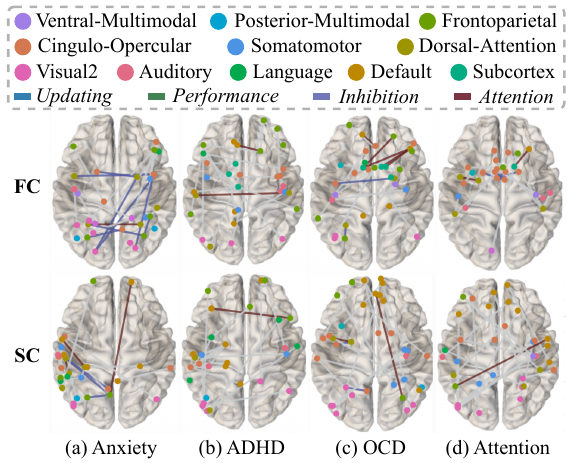}
    \caption{Top $0.01\%$ connections in the learned attention mask. Light-grey edges do not belong to any priors.}
    \label{fig:brain}
\end{figure}

To address \textbf{RQ5}, we visualize the top-weighted connections in averaged learned attention mask (Figure~\ref{fig:brain}). BrainTAP identifies critical connections primarily from the Inhibition and Attention priors. Attention is more prominent here than in Figure~\ref{fig:heatmap} because a small number of highly discriminative attention-related connections drive prediction, even though the prior ranks lower overall. Additionally, FC identifies more critical regions than SC, indicating it captures more diagnostically relevant information for psychiatric prediction.

\subsection{Conclusion}
In this paper, we present BrainTAP, a transformer framework that includes Adaptive Mutual Distillation (AMD) and Selective Prior Fusion (SPF) to predict brain disorders from FC and SC while preserving modality-specific signals and selectively integrating expert priors via global and personalized attention masks. Experiments on ABCD show that BrainTAP surpasses strong baselines. Further analyses reveal that AMD effectively regulates inter-modal interaction strength, and SPF highlights neurobiological circuits--particularly those involved in cognitive control and inhibition--as key discriminative features, aligning with known clinical evidence.

\noindent\textbf{Compliance with Ethical Standards.}
This study used openly available human data ABCD. Under the license, no additional institutional ethics approval was required.

\noindent\textbf{Acknowledgment.}
This work was supported by the UVA TYDE Seed Grant and the UVA Brain Institute Presidential Fellowship in Collaborative Neuroscience.

\bibliographystyle{IEEEbib}
\bibliography{strings,refs}

\end{document}